\documentclass[aps,pra,longbibliography,showkeys,showpacs,groupedaddress,twocolumn,10pt]{revtex4-1}
\usepackage{amssymb}
\usepackage{graphicx}
\usepackage{dcolumn}
\usepackage{bm}
\usepackage{amsmath}
\usepackage{subfigure}
\usepackage{float}
\usepackage{color}
\usepackage{float}
\usepackage{multirow}
\usepackage{diagbox}

\begin{document}
\title{Adiabatic potentials of cesium $(nD_J)_2$ Rydberg-Rydberg macrodimers}

\author{Xiaoxuan Han$^{1,2}$}
\author{Suying Bai$^{1,2}$}
\author{Yuechun Jiao$^{1,2}$}
\author{Georg Raithel$^{1,3}$}
\author{Jianming Zhao$^{1,2}$}
\thanks{Corresponding author: zhaojm@sxu.edu.cn}
\author{Suotang Jia$^{1,2}$}
\affiliation{$^{1}$State Key Laboratory of Quantum Optics and Quantum Optics Devices, Institute of Laser Spectroscopy, Shanxi University, Taiyuan 030006, China}
\affiliation{$^{2}$Collaborative Innovation Center of Extreme Optics, Shanxi University, Taiyuan 030006, China}
\affiliation{$^{3}$Department of Physics, University of Michigan, Ann Arbor, Michigan 48109-1120, USA}

\date{\today}

\begin{abstract}
Electrostatic multipole interactions generate long-range Rydberg-Rydberg macrodimer. We calculate the adiabatic potentials of cesium $(nD_{J})_2$ Rydberg macrodimers for principal quantum numbers $n$ ranging from 56 to 62, for $J=3/2$ and $5/2$, and for the allowed values of the conserved sum of the atomic angular-momentum components along the internuclear axis, $M$. For most combinations $(n, M, J)$ exactly one binding potential exists, which
should give rise to Rydberg macrodimer states. We study the dependence of the adiabatic potentials on the size of the two-body basis sets used in the calculation, and on the maximal order, $q_{max}$, of the multipole terms included in the calculation.
We determine the binding energies and lengths of the binding adiabatic potentials, and investigate their scaling behaviors as a function of the effective principal quantum number; these parameters are relevant to experimental preparation of Rydberg-atom macrodimers. Avoided crossings between adiabatic potentials affect the well shapes and the scaling behaviors differently in two distinct domains in $n$. We discuss an experimental scheme for preparing $(nD_J)_2$ Rydberg-atom macrodimers using two-color double-resonant photoassociation.

\end{abstract}
\pacs{32.80.Ee, 33.20.-t, 34.20.Cf}
\maketitle

\section{Introduction}
Rydberg atoms, highly excited states with large principal quantum number $n \gtrsim 10$, have been of interest in recent years due to their exaggerated properties~\cite{Gallagher}, for example large sizes and electric-dipole transition matrix elements ($\sim n^{2}$), and strong long-range van der Waals interactions ($\sim n^{11}$). The strong interaction between Rydberg atoms results in an excitation blockade effect~\cite{Comparat,Tong,Vogt2006,Vogt2007}, which has led to a variety of interesting investigations and applications, including quantum logic gates~\cite{Isenhower,J.yang2016}, single-photon sources~\cite{Dudin,Peyronel}, single-photon transistor~\cite{W.Li2014} and many-body systems and entanglement~\cite{Urban,A2009,Wilk}. Including a wider class of electric-multipole interactions in the analysis, one finds adiabatic potentials with minima that may support bound
Rydberg-Rydberg molecules~\cite{Boisseau,Overstreet,Deiglmayr2014,H2016,Han}. These molecules are macrodimers, as their bond lengths are $\thicksim$~4$n^2$ and can easily exceed 1~$\mu$m. Rydberg macrodimers have been predicted theoretically~\cite{Boisseau} and observed experimentally in cold atomic gases with  cesium~\cite{Overstreet,Deiglmayr2014,H2016,Han}, including $nD\,(n+2)D$ macrodimer~\cite{Overstreet}, $nS\, n'F$ and $nP \, nP$ molecules for 22$\leq n \leq 32$~\cite{Deiglmayr2014}, $43P \, 44S$~\cite{H2016} molecules bound by long-range dipolar interaction, and $(62D_J)_2$ Rydberg-atom macrodimers.
Certain Rydberg macrodimers are predicted to possess abundant vibrational states, large permanent dipole moments and exotic adiabatic potentials, which can be used to study vacuum fluctuations~\cite{L. H. Ford,G. Menezes}, quench ultracold collisions~\cite{Boisseau}, and measure correlations in quantum gases~\cite{Overstreet,M. Stecker}.

The adiabatic potentials play an important role for the preparation of Rydberg macrodimers.
The Rydberg macrodimers are identified by the assignment of photoassociation resonances to minima of the adiabatic potentials. In the present work, we investigate the adiabatic potentials of cesium Rydberg-atom pairs below the $(nD_J)_2$ ($n = 56-62$) asymptotes, calculated considering the electrostatic multipole interaction between Rydberg-atom pairs. We explore the effect of the basis size and the maximum interaction order on the potentials. We discuss the bonding energies of $(nD_J)_2$ Rydberg dimers, the corresponding equilibrium internuclear distances, and the effect of avoided potential crossings as a function of $n$.

\section{Multipole interaction Hamitonian}
For calculating the interaction of a Rydberg-atom pair, we consider two $nD_J$ Rydberg atoms, denoted $A$ and $B$, with an interatomic separation $\bf{R}$.
To simplify the calculation, the quantization axis and $\bf{R}$ are both chosen along the $z$-axis, see Fig.~1(a).
The relative positions of the Rydberg electrons of are ${\bf{r}}_A$ and ${\bf{r}}_B$. The interatomic distance $R$ is larger than the LeRoy radius~\cite{Le Roy}, $R_{LR}$, $i.e.$ the electronic wave functions do not overlap, and is small enough that radiation retardation effects~\cite{Casimir} are not important. The Hamiltonian of the Rydberg-atom pair is written as:

\begin{figure}[htbp]
\vspace{1ex}
\centering
\includegraphics[width=0.5\textwidth]{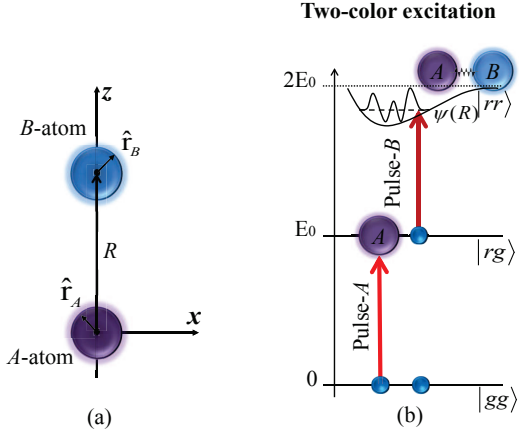}
\vspace{-1ex}
\caption{(a) Two-atom system. Rydberg atoms $\emph{A}$ and $\emph{B}$, separated by $R > R_{LR}$, are placed on the $z$-axis, ${\bf{r}}_{A}$ and ${\bf{r}}_{B}$ are the relative positions of the Rydberg electrons in atoms $\emph{A}$ and $\emph{B}$. (b) Level diagram and sketch of a vibrational wave-function for two-color double-resonant excitation of Rydberg-atom macrodimers. The pulse $A$ resonantly excites seed Rydberg atoms (atom $A$). The frequency of pulse $B$ is detuned relative to that of pulse $A$ by an amount equal to the molecular binding energy.}
\end{figure}

\vspace{-3ex}
\begin{eqnarray}
\hat{H} = \hat{H}_A + \hat{H}_B + \hat{V}_{int},
\end{eqnarray}
where $\hat{H}_{A(B)}$ is the Hamiltonian of atom $A(B)$, and $\hat{V}_{int}$ denotes the multipole interaction between the Rydberg-atom pair. $\hat{V}_{int}$ is taken as~\cite{Schwettmann,Deiglmayr2014,Han,Deiglmayr2016}

\begin{eqnarray}
\widehat{V}_{int}&=& \sum_{q=2}^{q_{max}} \frac{1}{R^{q+1}} \sum_{\substack{L_{A}=1 \\ L_B=q-L_A}}^{q_{max}-1}
\sum_{\Omega=-L_{<}}^{L_{<}}f_{AB\Omega}
\hat{Q}_{A} \hat{Q}_{B} \\
%\label{eq:Vin}
f_{AB\Omega}&=&\frac{(-1)^{L_{B}} \, (L_{A}+L_{B})!}{\sqrt{(L_{A}+\Omega)!(L_{A}-\Omega)!(L_{B}+\Omega)!(L_{B}-\Omega)!}}
\end{eqnarray}
where $L_{A(B)}$ are the multipole orders of atoms $\emph{A(B)}$, and the $L_{<}$ is the lesser of $L_{A}$ and $L_{B}$. The sum over $q$=$L_{A}$+$L_{B}$ starts at 2, because the atoms are neutral and have no monopole moment, and is truncated at a maximal order $q_{max}$. The factor $f_{AB\Omega}$ depends on $L_{A}$, $L_{B}$ and the counting index $\Omega$ under the third sum. The $\hat{Q}_{A(B)}$ are expressed as:

\begin{eqnarray}
\hat{Q}_{A}&=&\sqrt{\frac{4\pi}{2L_{A}+1}}\widehat{r}_{A}^{L_{A}}Y_{L_{A}}^{\Omega}(\widehat{\bf{r}}_{A})\\
\hat{Q}_{B}&=&\sqrt{\frac{4\pi}{2L_{B}+1}}\widehat{r}_{B}^{L_{B}}Y_{L_{B}}^{-\Omega}(\widehat{\bf{r}}_{B})
\end{eqnarray}
where the single-atom operators $\hat{{\bf{r}}}_{A}$ and $\hat{{\bf{r}}}_{B}$ are the relative positions of the Rydberg electrons in atoms $\emph{A}$ and $\emph{B}$,
the operators $\hat{Q}_{A(B)}$ include radial matrix elements, $\hat{r}_{A(B)}^{L_{A(B)}}$, and spherical harmonics that depend on the angular parts of the Rydberg-electron positions, $Y_{L_{A(B)}}^{\pm \Omega}(\hat{{\bf{r}}}_{A(B)})$.

We diagonalize the Hamiltonian of the Rydberg-atom pair on a dense grid of the internuclear separation, $R$. Considering global azimuthal symmetry, the projection of the sum of the electronic angular momenta, $M=m_{JA} + m_{JB}$, is conserved. For the homonuclear diatomic system in this work, the inversion symmetry is employed to define symmetrized basis states~\cite{Weber}, with $p=+1$ for even-parity states, $|\Psi_{g}\rangle$, and $p=-1$ for odd-parity states, $|\Psi_{u}\rangle$, that are not coupled by  $\hat{V}_{int}$.
In order to obtain the adiabatic potential curves of the Rydberg-atom pair, the Hamiltonian matrix is separately diagonalized for even- and odd-parity states using various  basis sizes (Sec.~\ref{sec:basis}) and values for the maximal multipole order, $q_{max}$ (Sec.~\ref{sec:qmax}), for a range of principal quantum numbers (Sec.~\ref{sec:nval}).

\section{Basis size dependence}
\label{sec:basis}

Numerical calculations of potential curves require a suitable  basis set, which, together with $q_{max}$, determines the number of interaction matrix elements used. The single-atom basis states are denoted $|n \ell J m_{J}\rangle$, and the corresponding Rydberg-atom pair basis states are denoted $|n_{A}\ell_{A}J_{A}m_{JA}\rangle$ $\otimes$ $|n_{B}\ell_{B}J_{B}m_{JB}\rangle$, with Rydberg atoms $A$ and $B$. For the quantum defects of the atomic energies~\cite{Seaton} and the fine structure coupling constants in $\hat{H}_A$ and $\hat{H}_B$ we use published values.
For high-$n$ ($n$=56-62 here), the atomic hyperfine coupling and the molecular rotation energy and are orders of magnitude smaller than the molecular interaction energy $\hat{V}_{int}$; they are therefore neglected. For the single-atom basis states, the range of effective principal quantum numbers is chosen as ${\rm{int}}(n_{\rm eff0})- \delta < n_{\rm eff} < {\rm{int}} (n_{\rm eff0})+\delta+1$, where $n_{\rm eff0}$ is the effective quantum number of the Rydberg state of interest, ${\rm{int}} (n_{\rm eff0})$ denotes the integer part of $n_{\rm eff0}$, and $\delta$ is a parameter for the principal quantum number range. Further, the single-atom orbital angular momentum space is restricted to $\ell \leq \ell_{max}$ and $m_{J} \leq m_{Jmax}$, for both atoms $A$ and $B$. The energies of the two-body molecular states are measured relative to a
state of interest, which in our case is of the type $n D_J n D_{J'}$, with $J, J' = 3/2$ or $5/2$. The two-body basis is then further truncated to two-body states with energy defects less than an upper limit of $25-30$~GHz from the molecular state of interest.

To test basis-size dependence, we calculate adiabatic potentials of 60$D_{5/2}$-atom pairs with two different basis sizes for $M = 0$ and $q_{max}$ = 6. One case is $54.9 < n_{\rm eff} < 60.1$ ($\delta=2.1$), and $\ell_{max}=4$, and $J, \vert m_J \vert \leq 4.5$, which results in a number of 4984 two-body basis states (= 2 $\times$ 2492 symmetrized two-body states), see Fig.~2(a). The other case is for $53.9 < n_{\rm eff} < 61.1$ ($\delta=3.1$), and $\ell_{max}= 5$, and $J, \vert m_J \vert \leq~5.5$, corresponding to a number of 11864 two-body basis states (=2$\times$ 5932 symmetrized two-body states). The calculated potential curves are shown in Fig.~2(b). For larger $\vert M \vert$-values, considered further below, the number of two-body states drops.

\begin{figure}[htbp]
\vspace{1ex}
\centering
\includegraphics[width=0.45\textwidth]{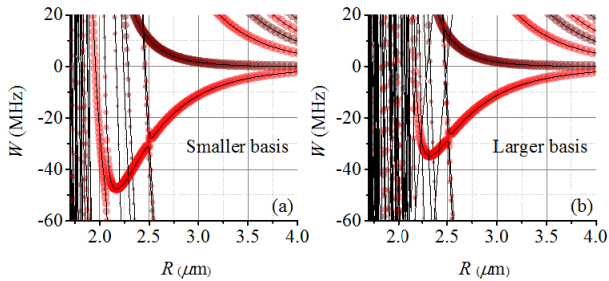}
\vspace{-1ex}
\caption{Calculations of adiabatic potential curves for cesium Rydberg-atom pairs, $(60D_{5/2})_2$,  with $M=0$ (black lines) and $q_{max}$ = 6 for a smaller (a) and a larger two-body basis size (b), with detailed truncation conditions explained in the text. Isotropic averages of laser excitation rates from the launch state $6P_{3/2} 60D_{5/2}$ are proportional to the areas of the overlaid circles (red for symmetric, p =+1, and deep-red for anti-symmetric states, p =-1).}
\end{figure}

Both in Figs. 2(a) and (b), most potential curves with small excitation rates do not exhibit minima that could give rise to bound macrodimer states, except one binding potential. In the range $R \lesssim 2.1~\mu$m, in Fig.~2(b), there are more potential curves, with some crossings, than in Fig.~2(a). However, these are not expected to produce observable effects, as there are no prominent wells associated with any of these steep, short-range potentials. In the range $R \gtrsim 2.1~\mu$m, the differences between the small- and large-basis calculations are more subtle. Both cases have one potential curve with a wide minimum conducive to bound molecular states.
Close inspection of Fig.~2(a) shows that the potential well with large oscillator strength exhibits a binding energy of $V_{min}$ $\approx$ $47.7$~MHz, at a binding length of $R_{equ}$ $\approx$ $2.17~\mu$m.
Increasing the basis size leads to small changes in these parameters. In Fig.~2(b), we find a binding  energy that is 15$\%$ less than in Fig.~2(a), equivalent to about 8~MHz, and a binding length that is 2$\%$ larger, corresponding to an increase of about 50~nm. These changes are attributed to level repulsion from additional levels in Fig.~2(b), which push the potential minimum up and out by some amount. The changes are large enough to become observable in an experiment.

In a second test, we have considered the case (62$D_{5/2})_2$ with $M$ = 3, for basis truncation parameters up to $54.9 \leq n_{eff}  \leq 64.9$, and $\ell_{max} = 7$, and $J$, $|m_J| \leq 7.5$, with a maximum two-body energy defect 30~GHz (basis size up to 2 $\times$ 4857 symmetrized two-body states). We have seen that the potential minima with large oscillator strengths shift by amounts up to about 10~MHz, and that the increase in $|m_J|$-range has the largest effect.

In the present work, we cannot make a definite statement about whether convergence has been reached in Fig.~2(b), because at this time it is not practical for us to increase the two-body basis size far beyond about 12000. Additional work on convergence could possibly be performed at lower principal quantum numbers, where basis sizes are generally smaller, or by implementing the calculations on a high-performance computing platform.
In the following sections, we use the same quantum-number range as in Fig.~2(a), unless noted otherwise.

\section{Dependence on multipole order}
\label{sec:qmax}

In Eq.~(2), the electrostatic multipole interactions  in $\hat{V}_{int}$ scale as 1/$R^{q+1}$ [outer sum in Eq.~(2)]. The multipole series proceeds through dipole-dipole ($dd$), dipole-quadrupole/quadrupole-dipole ($dq$/$qd$), dipole-octupole/octupole-dipole ($do$/$od$), quadrupole-quadrupole ($qq$), dipole-hexadecupole/hexadecupole-dipole ($dh$/$hd$) interactions, and so on, where the first (second) letter stands for atom $A$($B$). In our notation, the interaction series is terminated at the maximum order, $q_{max}$.
For example, for $q_{max} = 2$ only the dipole-dipole ($dd$) interaction is included. The case $q_{max} = 3$ includes $dd$, $dq$ and $qd$ interactions, $q_{max}=4$ includes $dd$, $do$, $od$, $dh$, $qq$, and $hd$ interactions, and so on.
Since our calculation is based on numerical diagonalization of the Hamiltonian in Eq.~(1), it includes the multipole interactions up to order $q_{max}$ in a non-perturbative fashion, in all orders perturbation theory. For instance, for $q_{max}$=2 the second-order dipole-dipole interaction, $i.e.$ the usual van der Waals interaction, is automatically included.

To show how the value of $q_{max}$ modifies the adiabatic potentials, in Fig.~3 we present the calculated adiabatic potential curves for the same quantum-number range as in Fig.~2(a), with $q_{max}$ =3 and 6.
Significant effects of interaction orders higher than $dd$ can be seen near the potential minima. In Fig.~3, the higher-order interactions push the potential well up by $\sim$6~MHz. Generally we have found that values of $q_{max}$ of 5 or 6 are sufficient to model the adiabatic potentials. Also, if the Rydberg-atom state pair has a strong resonant coupling to another state pair via an accidental F\"orster resonance of multipole order $q \leq 4$, at large distances $R$ the interaction energy scales as $1/R^{q+1}$, due to the $R-$scaling in Eq.~(2). If there is no low-order F\"orster resonance, at large distances the dipolar van der Waals interaction usually dominates (scaling $\propto 1/R^6$).

\begin{figure}[htbp]
\vspace{-1ex}
\centering
\includegraphics[width=0.45\textwidth]{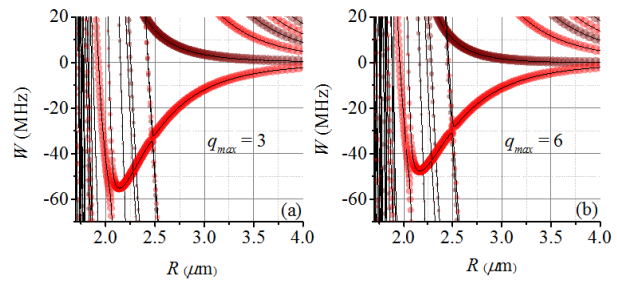}
\vspace{-1ex}
\caption{Adiabatic potential curves of 60$D_{5/2}$ Rydberg-atom pairs with the same quantum-number range as in Fig.~2(a) and $q_{max} = 3$ (a) and 6 (b).}
\end{figure}

The increase in computation time associated with an increase in $q_{max}$ was found to be very modest, in comparison with the increase as a function of the basis size. Therefore, we usually use $q_{max}=6$.

\section{Dependence on principal quantum number}
\label{sec:nval}

To determine the scaling behavior of the well depths and binding lengths with principal quantum number $n$, we have performed a series of calculations of the adiabatic potentials for $nD_J$ Rydberg-atom pairs with $n$ = 56 to 62, and $J$ = 3/2 or 5/2. The basis truncation parameters were $\delta$ =2.1, $\ell \leq 4$, and $J, \vert m_J \vert \leq 4.5$, and maximal multipole order $q_{max}$ = 6.
From the calculations, for each $n$ we have extracted the adiabatic potentials below the ($nD_{J})_2$ asymptotes that exhibit a binding potential well, which should give rise to Rydberg macrodimer states. For most combinations ($n, M, J$) exactly one such binding potential exists. For those, we determine the binding energies, $V_{min}$, and corresponding binding lengths, $R_{equ}$.
These parameters are relevant to the preparation of Rydberg-atom macrodimers in experiments.

In Fig.~4, we present $V_{min}$ and $R_{equ}$ for $nD_{5/2}$-atom pairs.
It is seen that the binding energy $|V_{min}|$ decreases and binding length $R_{equ}$ increases with increasing $n$. For all cases of $M$, $V_{min}$ exhibits a pronounced discontinuity between $n$=58 and 59, which will be discussed further below in Fig.~5. In the following, we explain the results for $M$ = 3 in detail; the results for all $M$ are then summarized in tables. Due to the discontinuity, we perform  partial allometric-function fits of the well depth over separate ranges $n$ = 56-58 and $n$=59-62,
\begin{eqnarray}
y = a(n-\delta_l)^b = a(n_{\rm eff0})^b \quad,
\end{eqnarray}
where $\delta_l$ =2.46 is the D-state quantum defect, $a$ and $b$ are the fitting parameters, with $b$ denoting the power scaling in $n_{\rm eff0}$.
For $M$ = 3, the fit parameters $b_{V_{min}}$ = -4.08$\pm$0.18 for $n$=56-58 and -4.76$\pm$0.52 for $n$=59-62.
The respective fit parameter for the binding length $R_{equ}$,  $b_{R_{equ}}$ = 2.31$\pm$0.03.
To compare the fit results for the various cases, in Table I we display the parameters $V_{min}$ and $R_{equ}$, and in Table II we list the corresponding fit  parameters $b$  for all combinations ($n, M, J$).
Averaging  over $M$, it is $b_{V_{min}}(ave)$ =-4.7 and $b_{R_{equ}}(ave)$ =2.5. Therefore, the potential depth $V_{min}$ scales faster than the Kepler frequency (which scales as $n^{-3}$), and  the binding length faster than the Rydberg-atom size (which scales as $n^2$).

\begin{table*}
    \caption{The binding energy, $V_{min}$ in MHz, and corresponding equilibrium internuclear distance, $R_{equ}$ in $\mu$m, of the adiabatic potentials for Rydberg-atom macrodimers, $(nD_{J})_2$ ($n=56-62$, $J =5/2,3/2$) and $M$ =0,1,2,3,4. }
    \begin{center}
\begin{tabular}{ |c|c|c|c|c|c|c|c|c|c|c|c|c|c|c|c|c|c|c| }
	\hline
	\multicolumn{1}{ |c| }{$nD_{J}$}& \multicolumn{10}{ |c| }{$(nD_{5/2})_2$}& \multicolumn{8}{ |c| }{$(nD_{3/2})_2$}\\
	\hline
	\multicolumn{1}{|c|}{\multirow{1}{*}{$M=$}}& \multicolumn{2}{c|}{0} &\multicolumn{2}{c|}{1}&\multicolumn{2}{c|}{2}&\multicolumn{2}{c|}{3}&\multicolumn{2}{c|}{4}& \multicolumn{2}{c|}{0} &\multicolumn{2}{c|}{1}&\multicolumn{2}{c|}{2}&\multicolumn{2}{c|}{3}\\
\hline
\multicolumn{1}{|c|}{\multirow{1}{*}{$n$}}  & $V_{min}$ &  $R$ & $V_{min}$ &  $R$  & $V_{min}$ &  $R$  & $V_{min}$ &  $R$ & $V_{min}$ &  $R$ & $V_{min}$ &  $R$ & $V_{min}$ &  $R$  & $V_{min}$ &  $R$  & $V_{min}$ &  $R$\\
   \hline
  56 & -159.5 & 1.81 & -170.4 & 1.76 & -153.8 & 1.79 & -126.0 & 1.82 & -177.9 & 1.53 & -159.5 & 1.69 &-149.9 & 1.68 & -110.5 & 1.67 & -9.3 & 1.73 \\
  \hline
  57 & -147.4 & 1.89 & -157.1 & 1.85 & -142.5 & 1.87 & -117.5 & 1.90 & -162.5 & 1.61 & -137.8 & 1.79 & -129.2 & 1.78 & -94.9 & 1.76 & -6.0 & 1.84 \\
  \hline
  58 & -135.2 & 1.98 & -143.6 & 1.93 & -130.7 & 1.96 & -108.4 & 1.99 & -144.5 & 1.69 & -127.6 & 1.87 & -119.5 & 1.86 & -87.6 & 1.84 & -4.8 & 1.94 \\
  \hline
  59 & -53.6 & 2.08 & -53.6 & 2.08 & -52.4 & 2.06 & -45.4 & 2.06 & -35.2 & 1.92 & -98.9 & 1.98 & -91.4 & 1.97 & -62.0 & 1.96 & -  & - \\
  \hline
  60 & -47.7 & 2.17 & -47.7 & 2.17 & -46.5 & 2.15 & -40.8 & 2.14 & -30.6 & 2.01 & -25.6 & 2.22 & -20.8 & 2.23 & -6.8 & 2.33 & - & -\\
  \hline
  61 & -44.8 & 2.26 & -44.8 & 2.26 & -43.6 & 2.25 & -38.2 & 2.24 & -28.3 & 2.09 & -23.4 & 2.32 & -18.8 & 2.33 & -5.6 & 2.44 & - & - \\
  \hline
  62 & -42.2 & 2.35 & -42.2 & 2.35 & -41.0 & 2.33 & -35.8 & 2.33 & -26.7 & 2.18 & -21.7 & 2.42 & -17.4 & 2.43 & -5.1 & 2.55 & - & -\\
  \hline
\end{tabular}
    \end{center}
\end{table*}

\begin{figure}[htbp]
\vspace{1ex}
\centering
\includegraphics[width=0.45\textwidth]{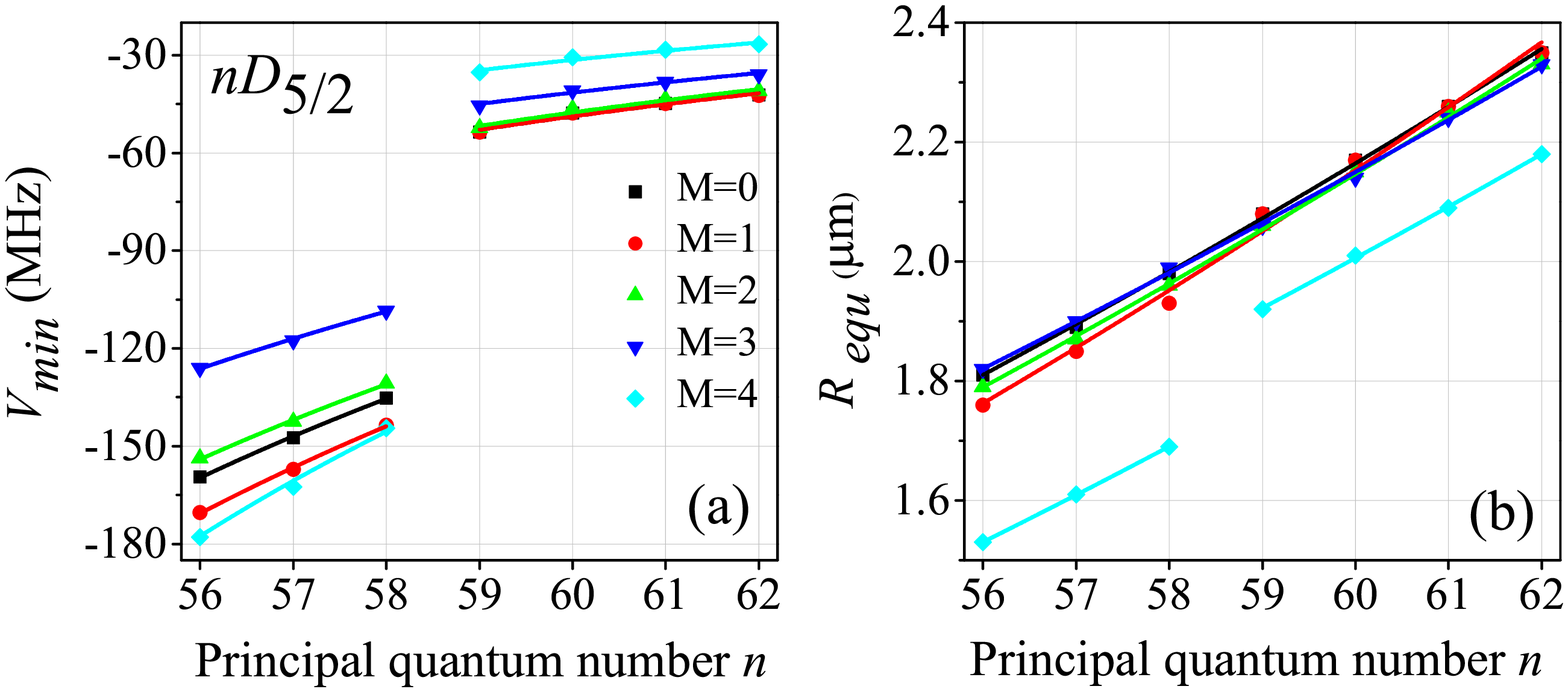}
\vspace{-1ex}
\caption{Calculations (symbols) of (a) binding energy, $V_{min}$, and (b) corresponding binding length, $R_{equ}$, of the adiabatic potential for cesium $(nD_{5/2})_2$ Rydberg macrodimers as a function of principal quantum number $n$. Solid lines show partial allometric fits, see text. }
\end{figure}

To  elucidate the origin of the discontinuities in Fig.~4, in Fig.~5 we plot the adiabatic potential curves for $M$ =3, for the cases $n$=59 (a) and $n$=58 (b). For $n \geq 59$, the potential exhibits one well, while for $n \leq 58$ the binding adiabatic potential is intersected by a repulsive potential that strongly couples with the binding potential, cutting the latter into two split wells. The two wells are separated by an avoided crossing that, in the case of Fig.~5(b), is centered at -120~MHz and has a gap size of about 40~MHz.
For $n \geq 59$, the avoided crossing moves out of the binding adiabatic potential, towards lower values of $R$, leaving merely a ledge [see circle in Fig.~5(a)].
Fig.~5 also shows how we define the potential depth $V_{min}$ plotted in Fig.~4 and tabulated in Table I, in the two domains of $n$. In the lower-$n$ domain, $n \leq 58$, the lower of the two split wells has a depth $\tilde{V}_{min}$ relative to the asymptotic level energy. As the lower well occurs at a smaller internuclear separation, it may be less likely to generate experimentally observable structures.

To complete our study of cesium $(nD_J)_2$ Rydberg molecules, we have performed calculations on $(nD_{3/2})_2$ molecules analogous to results discussed above for $J$=5/2. The data for $J$=3/2 are shown in Fig.~6 and are included in Tables I and II. As for $J$=5/2, in the case $J$=3/2 there exists one binding potential for most values of $M$ and $n$, and there is an avoided crossing that leads to a discontinuity of $V_{min}$ and $R_{equ}$ versus $n$ between $n$ =59 and 60, see Fig.~6. Overall, the potential depths and binding lengths are similar in the two cases of $J$. The binding length, $R_{equ}$, exhibits similar scaling in the two cases, see Table II. There is a notable difference in the power scaling of $V_{min}$; averaging over $M$ it is $V_{min}$ $\propto {n_{\rm eff0}}^{-8.44}$ for $n$ =56-59 and $\propto {n_{\rm eff0}}^{-6.28}$ for $n$ =60-62. These scalings are faster than for $J$=5/2. We attribute this difference to the $J$-dependence of level repulsion effects from nearby adiabatic potentials.

\begin{figure}[htbp]
\vspace{1ex}
\centering
\includegraphics[width=0.65\textwidth]{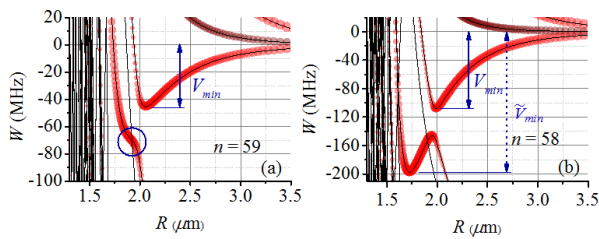}
\vspace{-1ex}
\caption{Adiabatic potentials for $(59D_{5/2})_2$ (a) and for $(58D_{5/2})_2$ (b) Rydberg pairs with $M$ =3. The truncation parameters are the same as in Fig.~2(a), and $q_{max}$=6.
The figure visualizes that an avoided crossing between two adiabatic potentials leads to a higher-$n$ domain with one binding well (a) and a lower-$n$ domain with two wells (b); more details see text.}
\end{figure}

\begin{figure}[htbp]
\vspace{1ex}
\centering
\includegraphics[width=0.45\textwidth]{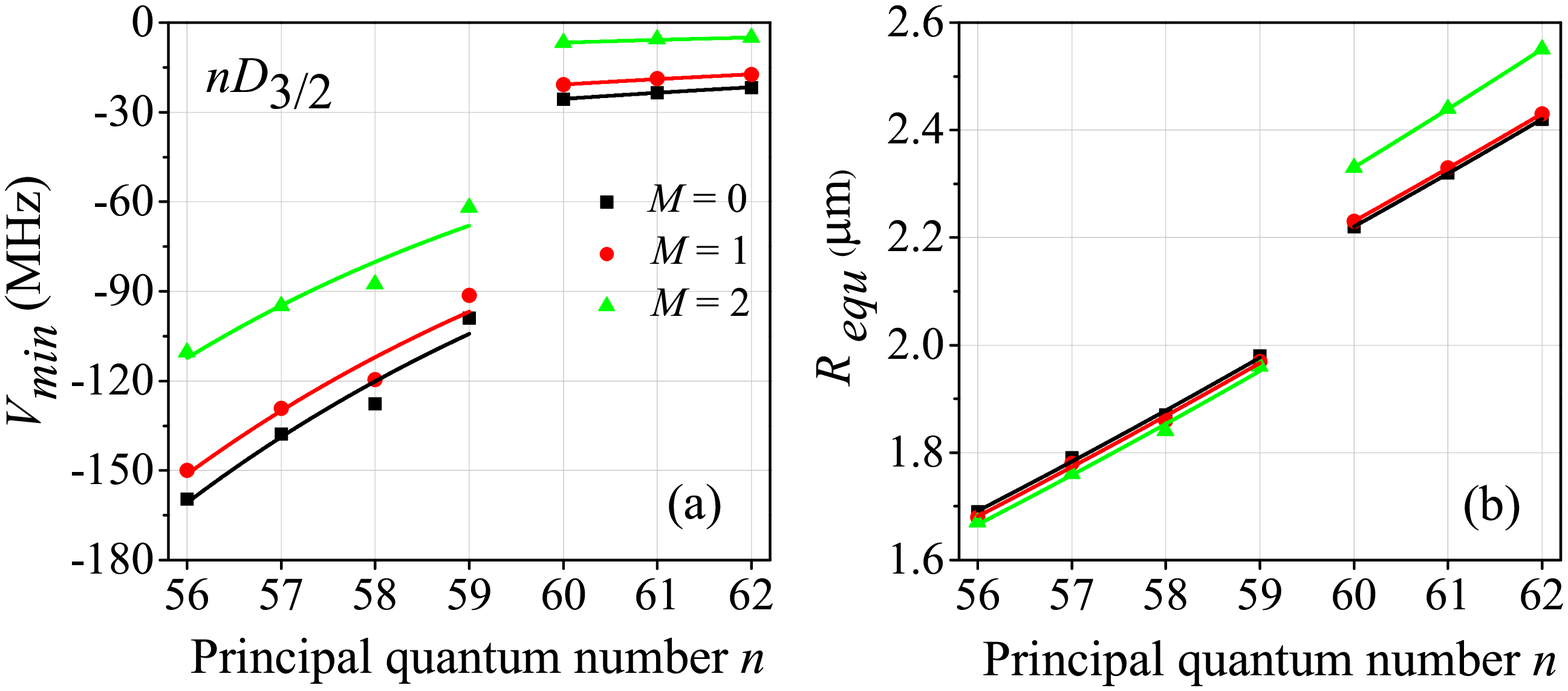}
\vspace{-1ex}
\caption{ Calculations (symbols) of binding energy, $V_{min}$, (a) and corresponding binding length, $R_{equ}$, (b) and fits (solid lines) for $(nD_{3/2})_2$ Rydberg macrodimers. The figure is analogous to Fig. 4. }
\end{figure}

\begin{table*}
    \caption{The fits parameters $b$ from Eq.~6 for the binding energy, $b_{V_{min}}$, and binding length, $b_{R_{equ}}$. Due to the discontinuities at $n$=58/59 ($J$=5/2) and at $n$=59/60 ($J$=3/2), we have performed separate fits for the respective low- and high-$n$ domains (see text). }
    \begin{center}
\begin{tabular}{ |c|c|c|c|c|c|c|c|c| }
	\hline
	\multicolumn{1}{|c|}{\multirow{1}{*}{$nD_{J}$}}& \multicolumn{4}{ |c| }{$(nD_{5/2})_2$}& \multicolumn{4}{ |c| }{($nD_{3/2})_2$} \\
    \hline
	\multicolumn{1}{|c|}{\multirow{1}{*}{$b$}}& \multicolumn{2}{c|}{$b_{V_{min}}$} &\multicolumn{2}{c|}{$b_{R_{equ}}$} &\multicolumn{2}{c|}{$b_{V_{min}}$} &\multicolumn{2}{c|}{$b_{R_{equ}}$}\\
	\hline
	\diagbox{$M$}{$n$}& \multicolumn{1}{c|}{$56-58$} &\multicolumn{1}{c|}{$59-62$}&\multicolumn{1}{c|}{$56-58$}&\multicolumn{1}{c|}{$59-62$}& \multicolumn{1}{c|}{$56-59$} &\multicolumn{1}{c|}{$60-62$}&\multicolumn{1}{c|}{$56-59$}&\multicolumn{1}{c|}{$60-62$}\\
\hline
  0 & -4.49$\pm$0.13 & -4.64$\pm$0.51 &\multicolumn{2}{c|}{2.48 $\pm$0.03} & -7.95$\pm$0.11 & -4.86$\pm$0.20 & 2.86$\pm$0.10 &2.52 $\pm$0.02  \\
  \hline
  1 & -4.65$\pm$0.15 & -4.60$\pm$0.52 &\multicolumn{2}{c|}{2.78 $\pm$0.10} & -8.18$\pm$1.15 & -5.26$\pm$0.34 & 2.87$\pm$0.10 & 2.51$\pm$0.02 \\
  \hline
  2 & -4.42$\pm$0.17 & -4.76$\pm$0.52 &\multicolumn{2}{c|}{2.52 $\pm$0.04} &-9.21 $\pm$1.63 & -8.70$\pm$1.38 & 2.90$\pm$0.15 & 2.64$\pm$0.02 \\
  \hline
  3 & -4.08$\pm$0.18 &-4.59 $\pm$0.37 &\multicolumn{2}{c|}{2.31 $\pm$0.03} & - & - & - & - \\
  \hline
  4 &-5.61 $\pm$0.40 & -5.45$\pm$0.68 &2.71$\pm$0.02 &2.43$\pm$0.04 &  - & - & - & - \\
  \hline
  $average$ & -4.65$\pm$0.57 & -4.81$\pm$0.36 &2.56$\pm$0.19 &2.50$\pm$0.17 & -8.44$\pm$0.67 & -6.28$\pm$2.11 & 2.88$\pm$0.02 & 2.56$\pm$0.07 \\
  \hline

\end{tabular}
    \end{center}
\end{table*}

\section{Experimental method for preparing $(nD_J)_2$ Rydberg-Rydberg macrodimers}

Based on the calculations above, we put forward an experimental proposal for preparing $(nD_J)_2$ Rydberg-Rydberg macrodimers using a two-color double-resonant photoassociation method. The level diagram and two-color excitation sketch are displayed in Fig.~1(b). The first color (laser pulse $A$) resonantly excites Rydberg atom-$A$ (seed atoms) from the ground state, and the second color (laser pulse $B$) is detuned relative to  pulse $A$ by an amount equal to the molecular binding energy. This sequence excites  $B$-Rydberg-atoms close to the $A$-atoms at a distance where a metastable Rydberg-Rydberg macrodimer exists. The frequency difference between the two colors yields the molecular bonding energy.

In practical implementations, the pulses $A$ and $B$ may involve more than one laser. For instance, for cesium one may use coincident 852-nm and 510-nm pulses that populate Rydberg levels through the intermediate $6P_{3/2}$ state. We have implemented such a scheme in our previous work~\cite{Han}.

\section{Conclusion}

We have numerically calculated the adiabatic potentials of cesium $nD_J$ Rydberg-atom pairs generated by electrostatic multipole interactions between Rydberg atoms.
We have tested the convergence of the results as a function of the basis size and maximal multipole interaction order. We have found that a maximum order of 6 is sufficient for convergence. However, for the high quantum numbers used in our work, it is not certain that basis sizes of about 12 thousand - the approximate limit in the present study - is sufficient to guarantee convergence; uncertainties of the potential depths on the order of 10~MHz may persist. We have determined the scaling behavior of the potential depth and binding length as function of the effective principal quantum number, $n_{\rm eff0}$, and discussed our findings. We have seen that avoided crossings between adiabatic potentials lead to the emergence of double-well structures, which cause discontinuities of the well depths as a function of $n_{\rm eff0}$. We also provide a two-color double-resonant photoassociation scheme for preparing $(nD_J)_2$ Rydberg-Rydberg macrodimers in experiments. Future work will deal with the convergence issue of Rydberg-pair molecule calculations, using larger product Hilbert spaces, as well as a survey of Rydberg-pair molecules with different angular-momentum quantum numbers (that may also be different for the two atoms involved). The results provide guidance for experimental work.

The work was supported by the National Key R$\&$D Program of China (Grant No. 2017YFA0304203), the National Natural Science Foundation of China (Grants No. 61475090, 61675123 and 61775124), Changjiang Scholars and Innovative Research Team in University of Ministry of Education of China (Grant No. IRT13076), and the State Key Program of National Natural Science of China (Grant No. 11434007). GR
acknowledges support by the National Science Foundation (PHY-1506093) and BAIREN plan of Shanxi province.

\end{document}